\documentclass{emulateapj}

\usepackage{natbib}

\usepackage{graphicx}
\usepackage[space]{grffile}
\usepackage{latexsym}
\usepackage{amsfonts,amsmath,amssymb}
\usepackage{url}
\usepackage[utf8]{inputenc}
\usepackage{fancyref}
\usepackage{hyperref}
\hypersetup{colorlinks=false,pdfborder={0 0 0},}
\usepackage{textcomp}
\usepackage{longtable}
\usepackage{multirow,booktabs}

\shorttitle{Disentangling Planets and Stellar Activity for Gliese 667C}
\shortauthors{Paul Robertson}

\begin{document}


\title{Disentangling Planets and Stellar Activity for Gliese 667C}


\author{Paul Robertson$^{1,2}$, Suvrath Mahadevan$^{1,2,3}$
\\
\normalsize{$^{1}$Department of Astronomy and Astrophysics, The Pennsylvania State University, pmr19@psu.edu}\\
\normalsize{$^{2}$Center for Exoplanets \& Habitable Worlds, The Pennsylvania State University}\\
\normalsize{$^{3}$The Penn State Astrobiology Research Center, The Pennsylvania State University}
}




\begin{abstract}
Gliese 667C is an M1.5V star with a multi-planet system, including
planet candidates in the habitable zone (HZ). The exact number of
planets in the system is unclear, because the existing radial velocity
(RV) measurements are known to contain contributions from stellar
magnetic activity. Following our analysis of Gliese 581 (Robertson et
al.~2014), we have analyzed the effect of stellar activity on the
HARPS/HARPS-TERRA RVs of GJ 667C, finding significant RV-activity
correlation when using the width (FWHM) of the HARPS cross-correlation
function to trace magnetic activity. When we correct for this
correlation, we confirm the detections of the previously-observed
planets b and c in the system, while simultaneously ascribing the RV
signal near 90 days (``planet d") to an artifact of the stellar
rotation. We are unable to confirm the existence of the additional RV
periodicities described in Anglada-Escud{{\'{e}}} et al.~(2013) in our
activity-corrected data.
\end{abstract}

\bibliographystyle{aj}

\section{Introduction}

Gliese 667C (GJ 667C), the tertiary companion to the K-dwarf binary GJ 667AB is known to host two super-Earth planets, based primarily on radial velocity (RV) observations from the ESO/HARPS spectrograph \citep{bonfils13,ae12,delfosse13}.  Of these planets--b ($P=7.2$d) and c ($P=28.1$d)--planet c is believed to orbit near the center of the circumstellar habitable zone \citep[HZ;][]{kopparapu13}, albeit on a significantly eccentric ($e=0.3$) orbit.  As many as five additional super-Earth planet candidates (GJ 667Cd,e,f,g,h, corresponding to periods of 92, 62, 39, 252, and 17 days) have been announced in the GJ 667C system by \citet{ae13}, using the HARPS-TERRA \citep[hereafter TERRA;][]{aebutler12} RV extraction pipeline.  However, determination of the number and nature of periodic RV signals for GJ 667C is complicated by the presence of RV contributions from stellar magnetic activity.  In particular, \citet{delfosse13} hesitated to claim any periodogram peaks between 90 and 372 days as planets since they appeared to be aliases of a single signal, with periodicity near 105 days (likely the stellar rotation period) detected in the changing width of the HARPS cross-correlation function (FWHM). \citet{makarov14} detect, in addition to the planets b and c, signals at 91.5, 53.3 and 35.3 days. They are careful to not claim these are planetary signals and note that the 53.3 and 35.3 are quite close to harmonics of a $\sim 106$ day rotation period.  Bayesian analysis of the velocities by \citet{feroz14} shows evidence for no more than three planets.  The current situation with Gliese 667C is therefore two detected signals widely accepted as exoplanets and additional periodicities whose exact significance and interpretation seem to differ among different groups.

We recently applied a simple, physically-motivated activity-RV correction for the multi-planet M dwarf system GJ 581 \citep{robertson14}, which was succesful in separating real planet signals from false positives created by activity.  Given the small number of exoplanet systems known via RV with multiple (super-)Earth-mass planets and/or HZ planets, it is essential to fully understand the effects of stellar activity on RV for these systems.  In this Letter we perform a similar activity analysis for GJ 667C.  Our analysis confirms the existence of planets b and c, while raising very significant doubts as to the planetary interpretation of the signals attributed to five super-Earths described in \citet{ae13}.

\section{Data}

We use 171 publicly available spectra of GJ 667C from the ESO HARPS archive\footnote{Based on data obtained from the ESO Science Archive Facility under request number 103301} for our analysis. While Keck HIRES and PFS RVS were also used by \citet{ae13}, the HARPS data formed the major portion of their RVs. To enable the most accurate analysis of both activity features and FWHM (the latter of which can only be accurately measured in the HARPS data) we restrict our analysis to HARPS spectra.  For each spectrum, we have computed the line strength indices for the magnetically-sensitive H$\alpha$ \citep[$I_{\textrm{H}\alpha}$;][]{robertson13} and Na I D \citep[$I_{\textrm{D}}$;][]{diaz07} lines.  We also consider the Ca II H\&K ($S_{HK}$), line bisector (BIS), and cross-correlation width (FWHM) activity tracers for these spectra.  BIS and FWHM values are computed by the HARPS data pipeline and included with the archived spectra \citep[as well as tabulated in][]{ae13}, and we adopt $S_{HK}$ values from \citet{ae13}.  With regards to BIS and FWHM, it is important to note that the intrinsic stability of the HARPS spectrograph ensures that any changes to the stellar line shapes reflected by changes in these indices are in fact due to stellar companions \citep{wright13} or stellar activity features (i.e. pulsations, spots, active regions) and not instrumental systematics \citep[see, e.g.][]{queloz09,santos10}. 

We have adopted the published RVs from the HARPS cross-correlation (hereafter HARPS) algorithm \citep[from][]{delfosse13} and the TERRA method \citep[from][]{ae13}, discussing differences between the techniques below.  We note here that because the statistical significance of \emph{all} observed periodic RV signals is higher using the TERRA RVs, we conclude these velocities are indeed more precise, and base all our final conclusions on this data set.

The HARPS spectra include one spectrum at $BJD=2454234.79$ which yields anomalously high values in every spectral activity index (potentially indicative of a flare), and another at $BJD=2454677.66$ with a FWHM that is $8.5\sigma$ higher than the rest of the data.  We exclude these spectra in our analysis.

\section{Analysis}

Our analysis follows from our recent examination of the activity-induced RV in GJ 581 \citep{robertson14}, and described in more detail therein.  Briefly, we search for a statistically significant correlation between RV and any of the available activity indicators.  We allow for the possibility of non-persistent starspots/active regions and amplitude/phase variability of the signals they induce by dividing the data into individual observing seasons.  The publicly available data include 6 spectra (4 in 2004-05 and 2 in 2009) for which the sampling is insufficient to evaluate activity-RV correlations.

We observe no activity-RV correlations prior to removing the RV signals of planet b and the linear acceleration caused by the star's orbit about the GJ 667AB binary.  However, upon removing these signals we observe statistically significant activity-RV correlations in the first (2006) and third (2008) full observing seasons.  These correlations are most significant for FWHM, with Pearson correlation coefficients of $r=0.62$ in 2006 and $r=0.55$ in 2008 for the TERRA RVs, yielding probabilities of no correlation $P(r)=5\times10^{-5}$, $4\times10^{-6}$, respectively.  These seasons show at least marginally significant activity-RV correlations in $I_{\textrm{D}}$ and $S_{HK}$ as well, with the exception of season 3 in $I_{\textrm{D}}$.  We note that when using the definition of $I_{\textrm{D}}$ of \citet{gds11}--who measure the emission in 0.5\AA~bands around the Na I D line centers (whereas we typically use 1\AA~windows)--$r$ increases from 0.17 to 0.31 for RV versus $I_{\textrm{D}}$ in season 3.  It is therefore possible that contamination in the line wings from the sodium sky emission lines or the GJ 667C photosphere is distorting the RV-$I_{\textrm{D}}$ correlation in season 3.  We list the full set of correlation values for the tracers in which we observed any significant activity-RV dependence in Table \ref{tab:actrv}.

We chose to correct for the effects of stellar activity on the RVs using the FWHM tracer because FWHM showed both the strongest correlations with RV and the most significant detection of the 105-day stellar rotation period \citep[in agreement with][]{delfosse13}. We computed linear least squares fits to the RV versus FWHM relations in the 2006 and 2008 seasons--with planet b and the linear trend removed--then subtracted those fits from the original RVs.  For the TERRA RVs, we find $RV = -1200~\textrm{m s}^{-1} + 0.39 \times FWHM$ in 2006 and $RV = -700~\textrm{m s}^{-1} + 0.20 \times FWHM$ in 2008, where RV and FWHM are in units of m s$^{-1}$.  Our $1\sigma$ uncertainties on the intercepts and slopes, respectively, are $300~\textrm{m s}^{-1}$ and 0.09 in 2006, $100~\textrm{m s}^{-1}$ and 0.04 in 2008. The two different fits accommodate changes in the activity level of the star, and its impact on RVs, in the different seasons. \\ \\

\begin{figure}[h!]
\centering
\includegraphics[width=0.95\columnwidth]{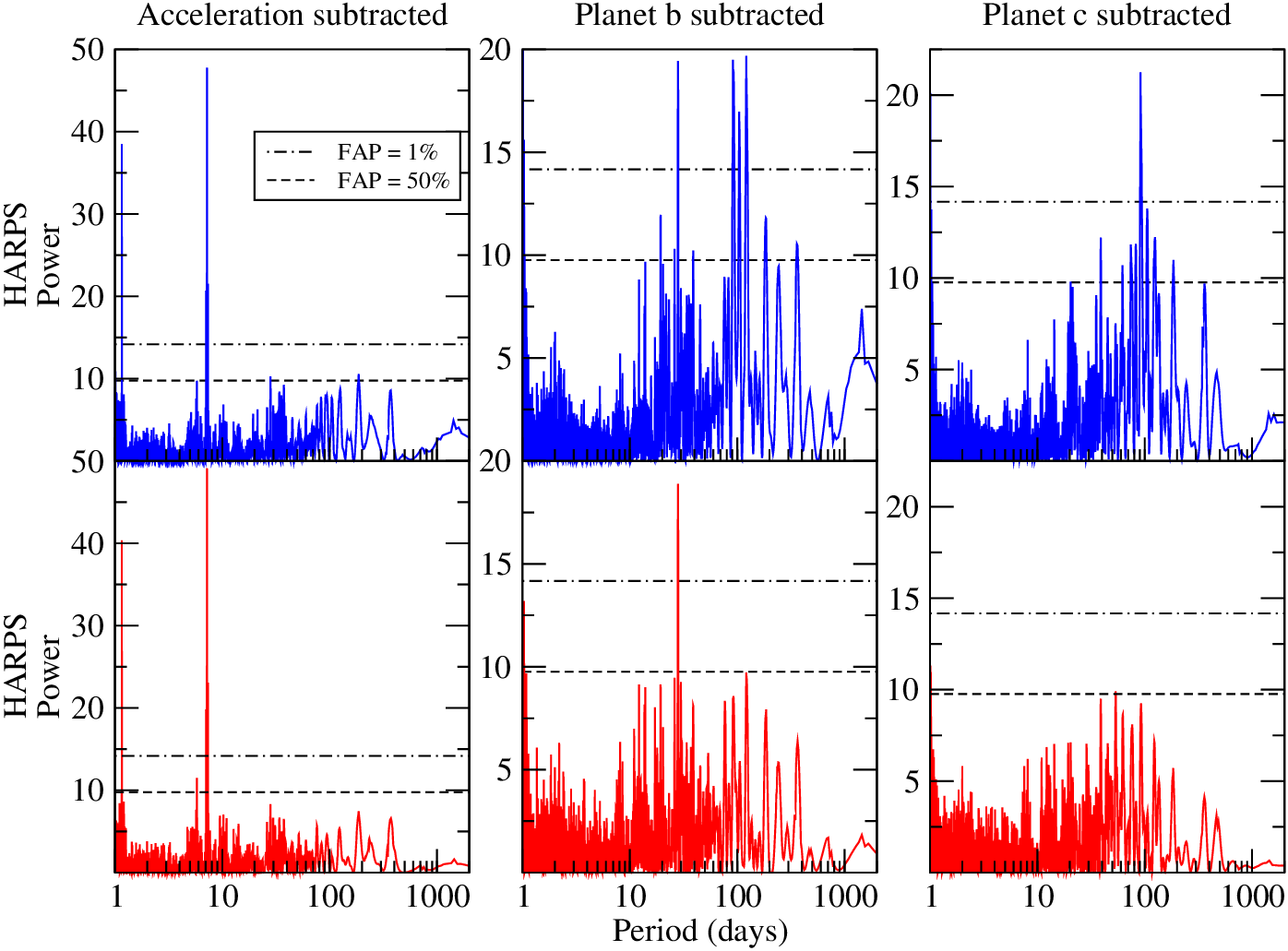}
\caption{\label{fig:rvps_harps}
Generalized Lomb-Scargle periodograms of the HARPS RVs before (\emph{blue}) and after (\emph{red}) correcting for the RV-FWHM activity correlation.  False-alarm probability (FAP) thresholds are given as dashed lines.}
\end{figure}

\begin{figure}[h!]
\centering
\includegraphics[width=0.95\columnwidth]{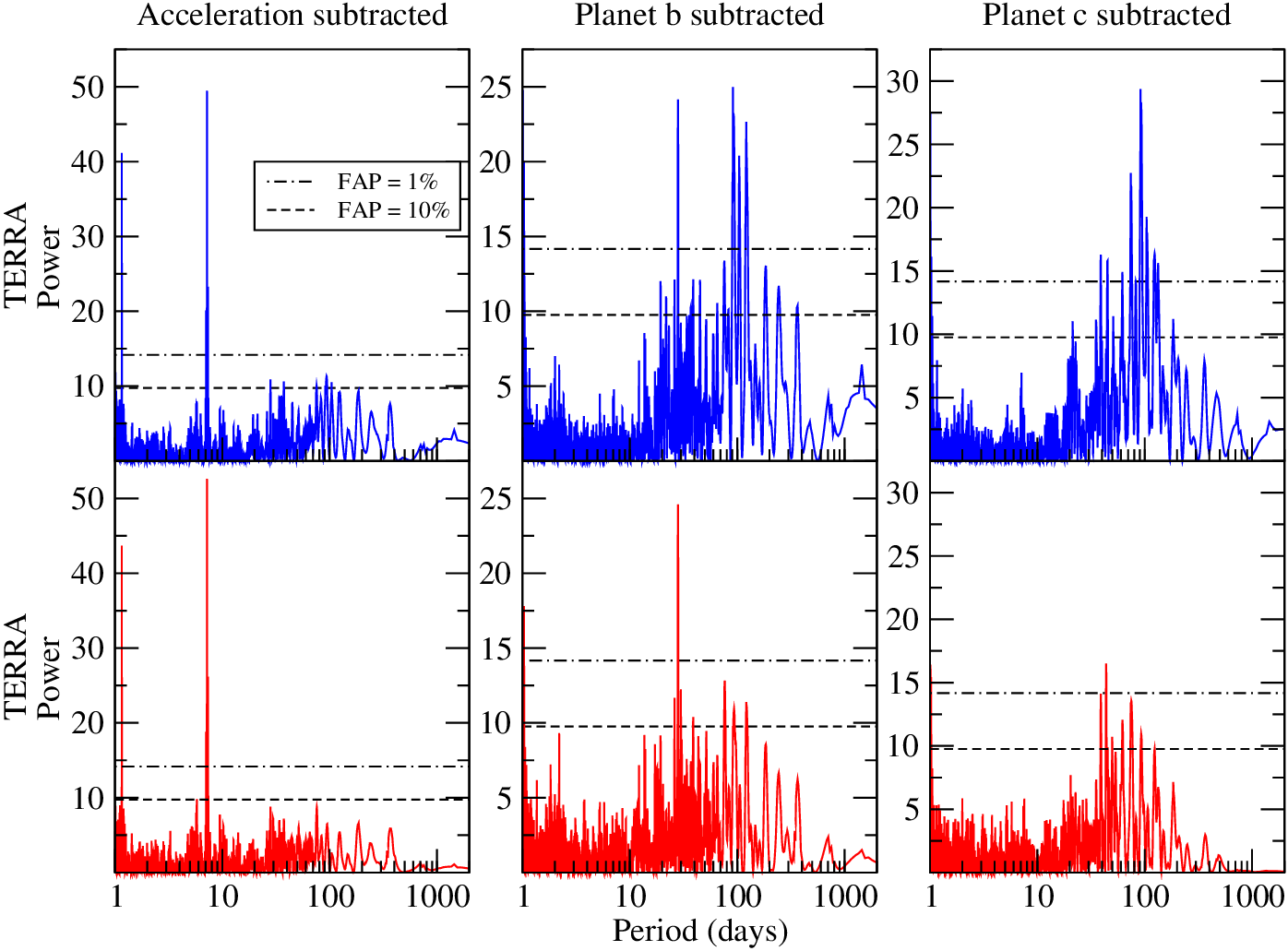}
\caption{\label{fig:rvps_terra}
Generalized Lomb-Scargle periodograms of the TERRA RVs before (\emph{blue}) and after (\emph{red}) correcting for the RV-FWHM activity correlation.  False-alarm probability (FAP) thresholds are given as dashed lines.}
\end{figure}

In Figures \ref{fig:rvps_harps} and \ref{fig:rvps_terra}, we show generalized Lomb-Scargle periodograms \citep{zk09} of the HARPS and TERRA RVs before and after applying our activity correction.  Accompanying power thresholds for false alarm probabilities (FAPs) of 1\% and 50\% are given according to Equation 24 of \citet{zk09}.  For both data sets, the peaks corresponding to planets b and c are approximately unchanged by the activity correction, indicating those planets are real.  On the other hand, the cluster of peaks centered on the 105-day rotation period is diminished significantly after correction, dropping below the 1$\sigma$ threshold for the HARPS RVs.  This suggests--as posited by \citet{delfosse13}--that these peaks do not correspond to planets, but are aliases of the stellar rotation signal.  It is worth noting that the star's X-ray luminosity $L_X = 26.87$ ergs\,s$^{-1}$ \citep{schmitt04} predicts a rotation period of $90 \pm 30$ days via the empirical age-rotation-$L_X$ relation of \citet{engle11}, so the rotation period obtained via FWHM is fully consistent with the star's X-ray emission.

We have modeled the orbits of planets b and c and the linear trend for this system using the activity-corrected RVs with the GaussFit \citep{jefferys88} and Systemic \citep{meschiari09} fitting routines, finding consistent results with both programs.  The orbital parameters are listed in Table 2.

The residual TERRA RVs around our orbital fit show marginally significant ($\sim3-4\sigma$) periodogram peaks near 44 and 75 days.  However, the significance of these peaks depends strongly on the parameters of the 2-planet fit, especially the eccentricity of planet c.  Furthermore, preliminary 3-planet fits with a planet at 44 days leads to solutions which are dynamically unstable within 100 years.  We therefore choose not to investigate these signals further, as we strongly suspect that--given their proximity to the original rotation signal--they are simply leftover power resulting from our imperfect activity-RV correction \citep[as seen for GJ 581;][]{robertson14}. \\ \\

\begin{figure}[h!]
\centering
\includegraphics[width=0.95\columnwidth]{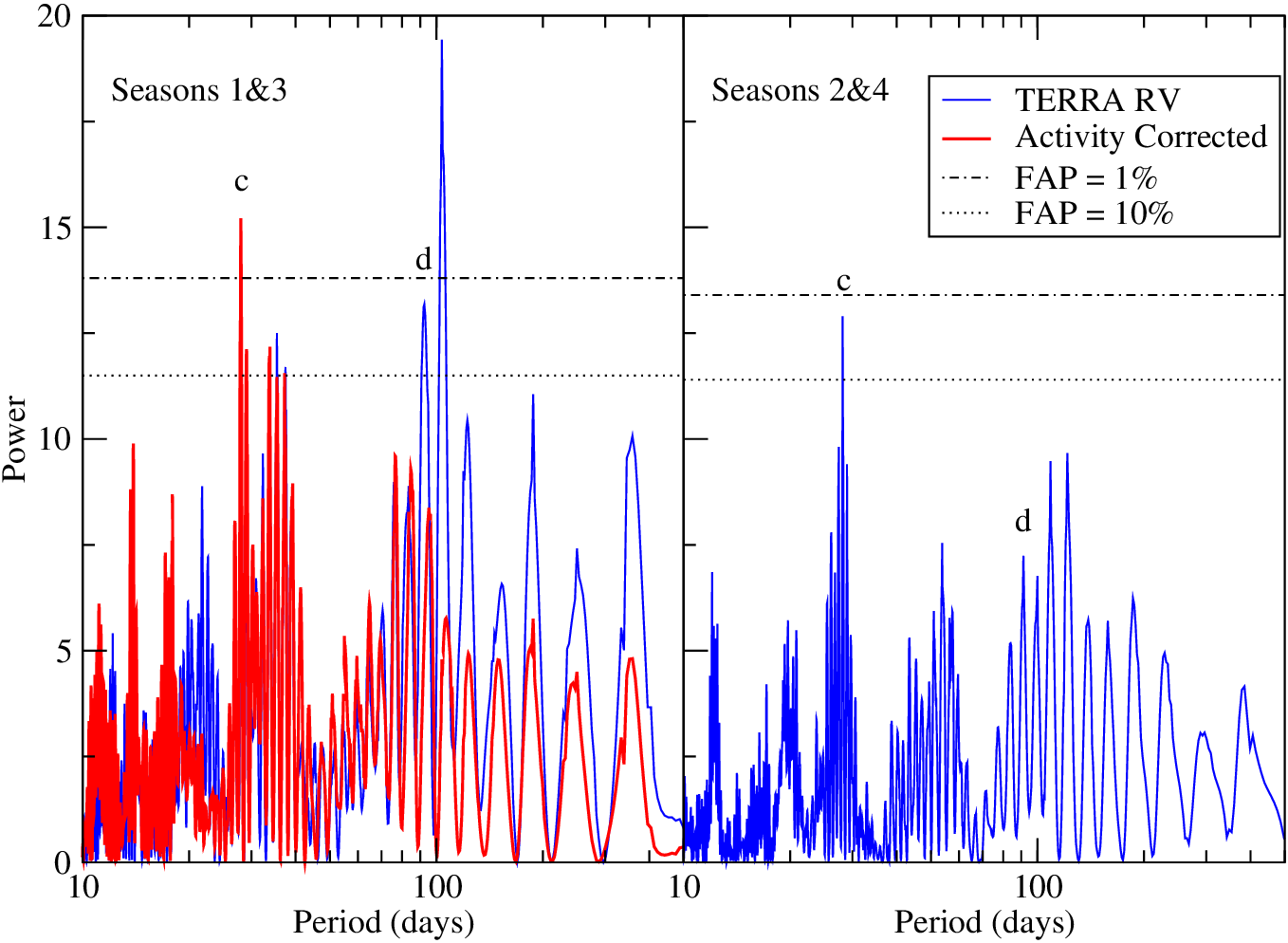}
\caption{\label{fig:seasons}
Comparison of periodograms for TERRA RVs separated by observing season.  Planet b and the linear acceleration have been subtracted.  The peaks corresponding to planet candidates c and d, and the thresholds for FAPs of 1\% and 10\% are indicated.}
\end{figure}

\section{Discussion}

The exquisite $\sim1$ m\,s$^{-1}$ precision afforded by the HARPS spectrograph on GJ 581 and GJ 667C has enabled detection of some of the lowest-mass exoplanets ever discovered with the Doppler radial velocity technqiue.  However, in these data sets we also see the need for thorough consideration of activity-RV correlations, as it appears activity-induced RV signals are ubiquitous at the m\,s$^{-1}$ level, even for quiet stars such as these.  We are encouraged, though, that our conceptually simple activity analysis has again clearly separated real exoplanet signals from false positives created by activity.

While our basic activity-RV algorithm has now provided promising results for two multi-planet systems (GJ 581 and GJ 667C), the results of this work show avenues for improvement.  Both $I_{\textrm{D}}$ and $S_{HK}$ showed correlations with RV, suggesting a ``master" activity indicator compiled from a number of tracers might offer a more complete description of the stellar activity.  A more sophisticated, flexible model for the functional activity-RV relation \citep[such as a Gaussian process model, see e.g.][]{roberts13,haywood14} may also improve our activity corrections.  These refinements will be the subject of future work.

The residuals of our 2-planet-plus-trend fit show no significant periodogram power at the periods of the ``planets" d ($P=92$d), e ($P=62$d), f ($P=39$d), or g ($P=256$d) proposed by \citet{ae13}.  In this regard, our conclusion is very similar to the Bayesian red noise analysis of \citet{feroz14}, with the important distinction that we are able to conclusively eliminate the 92-day signal as a stellar rotation alias using spectral diagnostics.  We note that \citet{ae13} considered RV correlations with activity indices--including  FWHM--but found results consistent with no correlation.  This discrepant result is most likely due to the fact that they did not separate the data into individual observing seasons to evaluate the activity dependence.  

While we show the 92-day signal is an activity signal, we cannot claim to have completely ruled out planets e-g.  These planet candidates were identified using a Bayesian likelihood analysis \citep[e.g.][]{tuomi13}, whereas we rely on the more traditional Lomb-Scargle periodogram to recover periodic signals.  It is therefore possible that the difference in analysis accounts for the non-detections; we are well aware that fitting for planets one at a time is suboptimal and can lead to decreased detection power \citep{aetuomi12,baluev14}.  However, given that our activity-RV correction algorithm tends to boost the signals of real low-mass planets \citep{robertson14}, we find the absence of periodogram power near the periods of these planet candidates to be compelling evidence against their existence. As we have already noted, other types of Bayesian analysis \citep{feroz14} do not find evidence for any additional planets. The Bayesian RV analysis must be repeated with a more careful treatment of stellar activity to convincingly demonstrate the existence of planets e-g.

\citet{ae13} partially justified the planetary nature of the GJ 667C planets by comparing periodograms in the first and second halves of the RV time series.  However, since the first and third (of four) observing seasons are affected by activity, each half of the series should be roughly equally contaminated by activity, leading to a false confirmation of ``planet d."  In Figure \ref{fig:seasons}, we show a modified version of this test, wherein we compare the periodogram of the first and third observing seasons (93 RV observations) with that of the second and fourth seasons (73 RV observations) after removing the linear trend and planet b.  Dividing the data this way, it is clear that the power near the period of the claimed planet d is not consistent from one half to the second.  In contrast, after correcting seasons 1 and 3 for activity, the power of planet c is quite similar in each half, further bolstering its status as a planet.

The results of this study imply an urgent mandate for the acquisition of additional data for GJ 667C.  In the case of GJ 581, time-series photometry revealed very low variability, indicating the magnetic activity creating the RV signals of false-positive planets d and g may be caused by localized active regions without optically dark starspots.  We are unaware of any such photometry for GJ 667C, and therefore cannot conclude whether the activity-induced RV shifts are caused by traditional starspots or spotless active regions.  Finally, more RV observations are required to determine whether additional low-mass planets truly exist in this system.

\begin{deluxetable*}{ccccccc}
\tablecaption{Activity-RV correlations}
\tablenum{1}
\startdata
\hline
\textbf{Season} &  \textbf{Dates} &  \textbf{$N_{RV}$} &  \textbf{$r_{FWHM}$} &  \textbf{$r_{I_{D}}$} &  \textbf{$r_{S_{HK}}$} &  \textbf{$P(r_{FWHM})$}  \\
1 &  February-September 2006 &  34 &  0.62 &  0.48 &  0.55 &  $5 \times 10^{-5}$ \\
2 &  March-September 2007 &  44 &  0.16 &  0.41 &  -0.03 &  $0.15$ \\
3 &  February-October 2008 &  59 &  0.55 &  0.17 &  0.34 &  $4 \times 10^{-6}$ \\
4 &  March-June 2010 &  27 &  -0.35 &  -0.23 &  -0.27 &  $0.04$  \\
\enddata
\tablecomments{Pearson correlation coefficients $r$ for the TERRA RVs as a function of several stellar activity indicators in each of four observing seasons for which HARPS spectra are publicly available.  The signal of GJ 667Cb and the linear acceleration have been subtracted from the RVs.  We include the probability of no correlation $P(r)$ between RV and FWHM, for which we see the strongest correlations.}
\label{tab:actrv}
\end{deluxetable*}

\begin{deluxetable*}{l|cc}
\tablecaption{Orbital model}
\tablenum{2}
\startdata
\hline
\textbf{Orbital Parameter} &     \textbf{Planet b} &  \textbf{Planet c} \\
Period $P$ (days)   & $7.1999 \pm 0.0009$   & $28.10 \pm 0.03$ \\
Periastron Passage $T_0$ & $4446.1 \pm 0.3$ & $4462 \pm 2$ \\
(BJD - 2 450 000) & &  \\
RV Amplitude $K$ (m/s) & $3.9 \pm 0.2$ & $1.9 \pm 0.3$ \\
Eccentricity $e$ & $0.15 \pm 0.05$ & $0.27 \pm 0.1$ \\
Longitude of Periastron $\omega$ & $12^{\circ} \pm 20^{\circ}$ & $140^{\circ} \pm 20^{\circ}$ \\
Semimajor Axis $a$ (AU) & $0.050431 \pm 0.000004$ & $0.12501 \pm 0.00009$ \\ 
Minimum Mass $M \sin i$ ($M_{\oplus}$) & $5.6 \pm 0.3$ & $4.1 \pm 0.6$ \\ 
\hline
Zero-point RV offset (m/s) & \multicolumn{2}{c}{$0.0 \pm 0.1$} \\
Linear Accceleration (m s$^{-1}$ yr$^{-1}$) & \multicolumn{2}{c}{$2.0 \pm 0.1$} \\
RMS (m/s) & \multicolumn{2}{c}{$1.84$} \\
\enddata
\tablecomments{Orbital solution for GJ 667C planets after correcting for stellar activity.  We use the HARPS-TERRA RVs to determine this solution, but note that the solution does not differ significantly when using the original HARPS RVs.}
\label{tab:orbits}
\end{deluxetable*}

\begin{acknowledgements}
We are grateful to the anonymous referee for a thoughtful and expeditious review.  We acknowledge support from NSF grants AST-1006676,  AST-1126413, AST-1310885, the Penn State Astrobiology Research Center, and the NASA Astrobiology Institute (NNA09DA76A) in our pursuit of precise RVs in the near infrared. This work was supported by funding from the Center for Exoplanets and Habitable Worlds. The Center for Exoplanets and Habitable Worlds is supported by the Pennsylvania State University, the Eberly College of Science, and the Pennsylvania Space Grant Consortium.
\end{acknowledgements}

\bibliography{gj667c_act_arxiv.bib}

\end{document}